\documentclass[12pt]{article}
\usepackage{bm,amsmath,amssymb,graphicx}

\begin{document}
\pagenumbering{arabic}
\begin{titlepage}

\title{The gravitational energy-momentum pseudo-tensor
of conformally invariant theories of gravity}

\author{F. F. Faria$\,^{*}$ \\
Centro de Ci\^encias da Natureza, \\
Universidade Estadual do Piau\'i, \\ 
64002-150 Teresina, PI, Brazil}

\date{}
\maketitle

\begin{abstract}
We construct the gravitational energy-momentum pseudo-tensor of up to 
fourth-order conformally invariant theories of gravity. Then we 
linearize the pseudo-tensor and use its average over a macroscopic 
region to find the energy and momentum carried by the 
plane gravitational waves of the three main conformally invariant theories 
of gravity.
\end{abstract}

\thispagestyle{empty}
\vfill
\bigskip
\noindent * felfrafar@hotmail.com \par
\end{titlepage}
\newpage


\section{Introduction}
\label{sec1}


It is well known that, according to the equivalence principle (EP), we can 
always nullify the gravitational field in an infinitesimal region of spacetime 
by a transformation of coordinates, meaning that we do not have a definition 
of local gravitational energy and momentum. However,  the theories of gravity 
that have the EP as a postulate, such as general relativity (GR), allows the 
definition of energy-momentum pseudo-tensors, which depend on the coordinate 
system, for the description of gravitational energy and momentum. Although 
the dependence on the coordinate system makes it impossible to obtain 
well-defined gravitational energy and momentum  from the pseudo-tensors, they 
are essential in the study of the energy and momentum carried by gravitational 
waves, which have been an important tool for understanding our 
universe since their first detection \cite{Ab}.

There are several definitions of gravitational energy-momentum pseudo-tensors
in GR \cite{Ein,Lan,Pap,Tol,Wei}, all of them leading to the same expression 
of the total energy of a system asymptotically flat obtained independently 
through the ADM canonical formulation\footnote{Some of these pseudo-tensors 
also give the same and reasonable result for various spacetimes 
in proper coordinates \cite{Vir1,Vir2,Vir3,Vir4,Vir5,Vir6,Vir7,Vir8}.}  
\cite{Arn}. 
However, such pseudo-tensors cannot be used in some alternative theories of 
gravity that have emerged to try to solve some GR problems, such as the well 
known dark matter \cite{Zwi} and cosmological constant \cite{Zel} problems, 
among others. Therefore, to check whether some of these alternative theories 
are consistent with observed gravitational wave data, it is necessary the 
development of gravitational energy-momentum pseudo-tensors for them. 

Recently, such pseudo-tensors have been developed for $f(R)$ \cite{Mul}, 
$n^{th}$-order \cite{Cap} and gauge theories of gravity \cite{Hob}. In this 
paper, we intend to develop a gravitational energy-momentum pseudo-tensor for 
up to fourth-order conformally invariant theories of gravity. In Sec. 2, we use 
a generalization of the method introduced by Bessel-Hagen for electromagnetism 
\cite{Bes} to derive a conformal gravitational energy-momentum pseudo-tensor. 
In Sec. 3, we linearize this pseudo-tensor and use its average over a region 
of spacetime to find the energy and momentum carried by the plane gravitational 
waves of the three main conformally invariant theories of gravity. Finally, 
in Sec. 4, we present our conclusions.


\section{The gravitational energy-momentum pseudo-tensor}
\label{sec2}


In general, the actions of up to fourth-order conformally invariant 
theories of gravity are of the form\footnote{By ``up to fourth-order'' 
we mean that the field equations of the theory have at most fourth 
derivatives of the fields.} 
\begin{equation}
S_g = \int{d^4x \, \mathcal{L}_g(g_{\mu\nu},\partial_{\lambda} g_{\mu\nu}, 
\partial_{\lambda}\partial_{\rho} g_{\mu\nu}, \varphi, 
\partial_{\lambda} \varphi, x^\mu)},
\label{1}
\end{equation}
where $\partial_{\mu}$ denotes ordinary derivatives, 
$g_{\mu\nu}$ is the metric tensor, $\varphi$ is a gravitational 
scalar field called dilaton, and $\mathcal{L}_g = \sqrt{-g}L_g$ is the 
gravitational Lagrangian density. It is worth noting that ({\ref{1}}) is 
invariant under both the coordinate transformations 
\begin{equation}
g'_{\mu\nu} = \frac{\partial x^{\alpha}}{\partial x'^{\mu}}
\frac{\partial x^{\beta}}{\partial x'^{\nu}} \, g_{\alpha\beta}, \ \ \ \ \ 
\varphi' = \varphi,
\label{2}
\end{equation}
and the (local) conformal transformations
\begin{equation}
\tilde{g}_{\mu\nu} = \Omega^{2}g_{\mu\nu}, \ \ \ \ \ \tilde{\varphi} = 
\Omega^{-1} \varphi,
\label{3}
\end{equation}
with $\Omega(x)$ being an arbitrary function of the spacetime coordinates.

According to the principle of least action $\delta S = 0$, the variation 
of (\ref{1}) with respect to $g_{\mu\nu}$, $\varphi$ and $x^\mu$ gives
\begin{eqnarray}
&& \int{d^4x} \Bigg[ \frac{\partial \mathcal{L}_g}{\partial 
g_{\alpha\beta}}\delta g_{\alpha \beta} + \frac{\partial \mathcal{L}_g}
{\partial(\partial_{\mu} g_{\alpha\beta})}\partial_{\mu}( \delta 
g_{\alpha \beta}) + \frac{\partial \mathcal{L}_g}{\partial(\partial_{\mu}
\partial_{\lambda} g_{\alpha\beta})}\partial_{\mu}\partial_{\lambda}(\delta 
g_{\alpha \beta}) \nonumber \\ && + \,  \frac{\partial \mathcal{L}_g}
{\partial \varphi}\delta \varphi + \frac{\partial \mathcal{L}_g}
{\partial(\partial_{\mu} \varphi)}\partial_{\mu}(\delta \varphi)
+ \partial_{\mu}\left(\mathcal{L}_g\delta x^{\mu}\right) \Bigg] = 0.
\label{4}
\end{eqnarray}
Integrating (\ref{4}) by parts, and neglecting surface terms, we obtain
\begin{equation}
 \int{d^4x} \, \left(\frac{\delta \mathcal{L}_g}{\delta{g_{\alpha\beta}}}
\delta g_{\alpha\beta} + \frac{\delta \mathcal{L}_g}{\delta{\varphi}} 
\delta\varphi+ \partial_{\mu}J^\mu \right)= 0,
\label{5}
\end{equation}
where
\begin{eqnarray}
\frac{\delta \mathcal{L}_g}{\delta{g_{\alpha\beta}}} &=& \frac{\partial 
\mathcal{L}_g}{\partial g_{\alpha\beta}} - \partial_{\mu}\left[ \frac{\partial 
\mathcal{L}_g}{\partial(\partial_{\mu} g_{\alpha\beta})}\right] 
+ \partial_{\mu}\partial_{\lambda}\left[ \frac{\partial \mathcal{L}_g}
{\partial(\partial_{\mu}\partial_{\lambda} g_{\alpha\beta})}\right],
\label{6} \\
\frac{\delta \mathcal{L}_g}{\delta{\varphi}} &=& \frac{\partial \mathcal{L}_g}
{\partial \varphi} - \partial_{\mu}\left[ \frac{\partial \mathcal{L}_g}
{\partial(\partial_{\mu} \varphi)}\right]
\label{7}
\end{eqnarray}
are the Euler–Lagrange variational derivatives, and
\begin{eqnarray}
J^{\mu} &=&  \left[\frac{\partial \mathcal{L}_g}
{\partial(\partial_{\mu} g_{\alpha\beta})} - \partial_{\lambda}
\left(\frac{\partial \mathcal{L}_g}{\partial(\partial_{\mu}\partial_{\lambda} 
g_{\alpha\beta})}\right) \right]\delta g_{\alpha\beta} 
+ \frac{\partial \mathcal{L}_g}{\partial(\partial_{\mu} \, \partial_{\lambda} 
g_{\alpha\beta})} \partial_{\lambda}(\delta g_{\alpha\beta}) \nonumber \\ && 
+ \, \frac{\partial \mathcal{L}_g}{\partial(\partial_{\mu} \varphi)}\delta \varphi
+ \mathcal{L}_g\delta x^{\mu}
\label{8}
\end{eqnarray}
is the Noether current.

By performing an infinitesimal coordinate transformation
\begin{equation}
\delta x^{\mu} = - \, \epsilon^{\mu},
\label{9}
\end{equation}
and an infinitesimal conformal transformation, on $g_{\mu\nu}$ and $\varphi$, 
we find
\begin{eqnarray}
\delta g_{\mu\nu} &=& \nabla_{\mu}\epsilon_{\nu} + \nabla_{\nu}\epsilon_{\mu} 
+ 2\omega g_{\mu\nu},
\label{10} \\
\delta \varphi &=&  \epsilon^{\mu}\nabla_{\mu}\varphi - \omega \varphi,
\label{11}
\end{eqnarray}
where $\nabla_{\mu}$ denotes covariant derivatives, $\epsilon^{\mu}$ is an 
arbitrary spacetime dependent vector field and $\omega$ is an arbitrary 
spacetime dependent scalar field. For $\epsilon^{\mu}$ to be
conformal, it must obey the conformal Killing equation
\begin{equation}
\nabla_{\mu}\epsilon_{\nu} + \nabla_{\nu}\epsilon_{\mu} = 2\omega g_{\mu\nu},
\label{12}
\end{equation}
from which we get
\begin{equation}
\omega = \frac{1}{4}\nabla_{\mu}\epsilon^{\mu}.
\label{13}
\end{equation}
The substitution of (\ref{13}) into (\ref{10}) and (\ref{11}) then gives
\begin{eqnarray}
\delta g_{\mu\nu} &=& \nabla_{\mu}\epsilon_{\nu} + \nabla_{\nu}\epsilon_{\mu} 
+ \frac{1}{2} g_{\mu\nu}\nabla_{\lambda}\epsilon^{\lambda},
\label{14} \\
\delta \varphi &=&  \epsilon^{\mu}\nabla_{\mu}\varphi -  \frac{1}{4}\varphi
\nabla_{\mu}\epsilon^{\mu},
\label{15}
\end{eqnarray}
which, after some calculation, becomes
\begin{eqnarray}
\delta g_{\mu\nu} &=& \left(\partial_{\lambda}g_{\mu\nu} 
- \frac{1}{4} g_{\mu\nu} g^{\rho\sigma}\partial_{\lambda}g_{\rho\sigma}\right)
\epsilon^{\lambda} + g_{\mu\lambda}\partial_{\nu}\epsilon^{\lambda} 
+ g_{\nu\lambda}\partial_{\mu}\epsilon^{\lambda}
+\frac{1}{2}g_{\mu\nu}\partial_{\lambda}\epsilon^{\lambda},
\label{16} \nonumber \\ && \\
\delta \varphi &=&  \left(\partial_{\lambda}\varphi
+ \frac{1}{8} \varphi g^{\rho\sigma}\partial_{\lambda}g_{\rho\sigma}\right)
\epsilon^{\lambda} - \frac{1}{4}\varphi \partial_{\lambda}\epsilon^{\lambda}.
\label{17}
\end{eqnarray}

Finally, substituting (\ref{16}) and (\ref{17}) into (\ref{5}), and using the 
fact that $\epsilon^{\mu}$ is global ($\partial_{\nu}\epsilon^{\mu}=0$), 
we obtain
\begin{eqnarray}
 &&\int{d^4x} \, \bigg[\frac{\delta \mathcal{L}_g}{\delta{g_{\alpha\beta}}}
\left(\partial_{\nu}g_{\alpha\beta} - \frac{1}{4} g_{\alpha\beta} 
g^{\rho\sigma}\partial_{\nu}g_{\rho\sigma}\right)
+ \frac{\delta \mathcal{L}_g}{\delta{\varphi}}\left(\partial_{\nu}\varphi
+ \frac{1}{8} \varphi g^{\rho\sigma}\partial_{\nu}g_{\rho\sigma}\right)
\nonumber \\ && \, + \frac{1}{c}\partial_{\mu}\left(\sqrt{-g}t^{\mu}\,\!_{\nu}\right) 
\bigg]\epsilon^{\nu}= 0,
\label{18}
\end{eqnarray}
where
\begin{eqnarray}
t^{\mu}\,\!_{\nu} &=& \frac{c}{\sqrt{-g}} \Bigg\{ 
\left[ \frac{\partial \mathcal{L}_g}{\partial(\partial_{\mu} g_{\alpha\beta})} 
-  \partial_{\lambda}\left(\frac{\partial \mathcal{L}_g}{\partial
(\partial_{\mu}\partial_{\lambda} g_{\alpha\beta})}\right)  \right]
\left(\partial_{\nu}g_{\alpha\beta} - \frac{1}{4} g_{\alpha\beta} 
g^{\rho\sigma}\partial_{\nu}g_{\rho\sigma}\right) \nonumber \\ && 
 + \, \frac{\partial \mathcal{L}_g}{\partial(\partial_{\mu} \, \partial_{\lambda} 
g_{\alpha\beta})} \partial_{\lambda}\left(\partial_{\nu}g_{\alpha\beta} 
- \frac{1}{4} g_{\alpha\beta} g^{\rho\sigma}\partial_{\nu}g_{\rho\sigma}\right)
 \nonumber \\ && 
+ \, \frac{\partial \mathcal{L}_g}{\partial(\partial_{\mu} \varphi)} 
\left(\partial_{\nu}\varphi + \frac{1}{8} \varphi g^{\rho\sigma}
\partial_{\nu}g_{\rho\sigma}\right)
- \delta^{\mu}\,\!_{\nu}\mathcal{L}_g\Bigg\}
\label{19}
\end{eqnarray}
is the gravitational energy-momentum pseudo-tensor\footnote{The dependence of 
$t^{\mu}\,\!_{\nu}$ on the ordinary derivatives of the metric tensor 
$g_{\mu\nu}$ causes it to not covariantly transform like a tensor under 
coordinate transformations.} of up to forth-order conformally invariant 
theories of gravity, with $c$ being the speed of light in vacuum. Using (\ref{3}), 
it can be shown that (\ref{19}) is covariant under conformal transformations 
and, consequently, it is traceless ($t = t^{\mu}\,\!_{\mu} 
= g^{\mu\nu}t_{\mu\nu} = 0$).

In the presence of conformally invariant matter, we have the Euler-Lagrange 
field equations
\begin{eqnarray}
\frac{\delta \mathcal{L}_g}{\delta{g_{\alpha\beta}}} &=&  
\frac{1}{2c}\sqrt{-g}T_{\alpha\beta},
\label{20} \\
\frac{\delta \mathcal{L}_g}{\delta{\varphi}} &=& 
- \frac{1}{\varphi c}\sqrt{-g}\left(T^{\mu}\,\!_{\mu}\right) = 0,
\label{21}
\end{eqnarray}
where 
\begin{equation}
T^{\mu\nu} = - \frac{2}{\sqrt{-g}} \frac{\delta \mathcal{L}_{m}}
{\delta g_{\mu\nu}}
\label{22}
\end{equation}
is the matter energy-momentum tensor, which is traceless due to the conformal 
symmetry of the matter fields, and we used the conformal Ward identity
\begin{equation}
2g_{\mu\nu}\frac{\delta\mathcal{L}_g}{\delta g_{\mu\nu}} 
+ \varphi\frac{\delta\mathcal{L}_g}{\delta \varphi} = 0
\label{23}
\end{equation}
in (\ref{21}), with $\mathcal{L}_{m} = \sqrt{-g}L_m$ being the Lagrangian 
density of the matter fields.

The use of (\ref{20}) and (\ref{21}) in (\ref{18}) gives
\begin{equation}
\int{d^4x} \,\left[ \frac{1}{2}\sqrt{-g}T^{\alpha\beta}
\left(\partial_{\nu}g_{\alpha\beta} 
- \frac{1}{4} g_{\alpha\beta} g^{\rho\sigma}\partial_{\nu}g_{\rho\sigma}\right)
+ \partial_{\mu}\left(\sqrt{-g}t^{\mu}\,\!_{\nu}\right)\right]\epsilon^{\nu} = 0.
\label{25}
\end{equation}
Considering that $T^{\mu\nu}$ is traceless and obeys the conservation law
\begin{equation}
\nabla_{\mu}T^{\mu}\,\!_{\nu} = \frac{1}{\sqrt{-g}}\partial_{\mu}
\left(\sqrt{-g}T^{\mu}\,\!_{\nu}\right) - \frac{1}{2}T^{\alpha\beta}\partial_{\nu}
g_{\alpha\beta} = 0,
\label{26}
\end{equation}
we can write (\ref{25}) as
\begin{equation}
\int{d^4x} \,\partial_{\mu}\left[\sqrt{-g}\left(T^{\mu}\,\!_{\nu} 
+ t^{\mu}\,\!_{\nu}\right)\right]\epsilon^{\nu} = 0.
\label{27}
\end{equation}
Since $\epsilon^{\mu}$ is arbitrary, it follows from (\ref{27}) that
\begin{equation}
\partial_{\mu}\left[\sqrt{-g}\left(T^{\mu}\,\!_{\nu} 
+ t^{\mu}\,\!_{\nu}\right)\right] = 0,
\label{28}
\end{equation}
which implies the conservation of the energy and momentum of both matter 
and gravity.

We can derive from (\ref{28}), the total four-momentum of matter 
plus gravity
\begin{equation}
P^{\mu} = \frac{1}{c}\int dx^3 \sqrt{-g}\left(T^{\mu 0}
+ t^{\mu 0}\right),
\label{29}
\end{equation}
and the total angular momentum of matter plus gravity
\begin{equation}
M^{\mu\nu} = \frac{1}{c} \int dx^3 \sqrt{-g}\left[x^\mu\left(T^{\nu 0}
+ t^{\nu 0}\right) - x^\nu\left(T^{\mu 0}
+ t^{\mu 0}\right)\right].
\label{30}
\end{equation}
Due to the non-covariance of $t^{\mu \nu}$ under coordinate transformations, 
we cannot use (\ref{29}) and (\ref{30}) to calculate the total energy, momentum 
and angular momentum in a finite region of space. However, these quantities can 
be used over the entire space for asymptotically flat spacetimes, since 
$t^{\mu \nu}$ transforms covariantly like a tensor under linear affine 
transformations.


\section{Linearized gravitational energy-momentum pseudo-tensor}
\label{sec3}


By performing the flat background field expansions
\begin{eqnarray}
g_{\mu\nu} &=&  \eta_{\mu\nu} + h_{\mu\nu},
\label{31} \\
\varphi &=& \varphi_0 + \sigma,
\label{32}
\end{eqnarray}
and keeping only the terms of second order in the infinitesimal perturbations 
$h^{\mu\nu}$ and $\sigma$, we find that (\ref{19}) reduces to the 
linearized gravitational energy-momentum pseudo-tensor
\begin{eqnarray}
\bar{t}^{\mu}\,\!_{\nu} &=& c \, \Bigg\{\left[ \frac{\partial 
\bar{\mathcal{L}_g}}{\partial(\partial_{\mu} 
h_{\alpha\beta})} - \partial_{\lambda}
\left(\frac{\partial \bar{\mathcal{L}_g}}{\partial(\partial_{\mu}
\partial_{\lambda} h_{\alpha\beta})}\right) \right]
\left(\partial_{\nu}h_{\alpha\beta} 
- \frac{1}{4} \eta_{\alpha\beta}\partial_{\nu}h \right) 
\nonumber \\ && 
 + \, \frac{\partial \bar{\mathcal{L}_g}}{\partial(\partial_{\mu} \, 
\partial_{\lambda} h_{\alpha\beta})} \partial_{\lambda}
\left(\partial_{\nu}h_{\alpha\beta} 
- \frac{1}{4} \eta_{\alpha\beta}\partial_{\nu}h \right)
 \nonumber \\ && 
+ \, \frac{\partial \bar{\mathcal{L}_g}}{\partial(\partial_{\mu} \sigma)} 
\left(\partial_{\nu}\sigma + \frac{1}{8} \varphi_0
\partial_{\nu}h\right)
- \delta^{\mu}\,\!_{\nu}\bar{\mathcal{L}_g}\Bigg\},
\label{33}
\end{eqnarray}
where $\eta_{\mu\nu} = \mbox{diag}(-1,+1,+1,+1)$ is the flat Minkowski 
metric, $\varphi_{0}$ is a constant background dilaton field, 
$\bar{\mathcal{L}_g}$ is the linearized gravitational Lagrangian density, and 
$h = h^{\mu}\,\!_{\mu} = \eta^{\mu\nu}h_{\mu\nu}$.

Taking the trace of (\ref{33}), considering that $\bar{t}^{\mu}\,\!_{\mu} = 0$, 
putting the result back into the average of (\ref{33}), and integrating by parts, 
we obtain
\begin{eqnarray}
\left\langle \bar{t}^{\mu}\,\!_{\nu}\right\rangle &=& c \, 
\bigg\langle\left[ \frac{\partial 
\bar{\mathcal{L}_g}}{\partial(\partial_{\mu} 
h_{\alpha\beta})} - 2 \partial_{\lambda}
\left(\frac{\partial \bar{\mathcal{L}_g}}{\partial(\partial_{\mu}
\partial_{\lambda} h_{\alpha\beta})}\right) \right]
\left(\partial_{\nu}h_{\alpha\beta} 
- \frac{1}{4} \eta_{\alpha\beta}\partial_{\nu}h \right) 
\nonumber \\ && 
-\frac{1}{4}\delta^{\mu}\,\!_{\nu}\left[ \frac{\partial 
\bar{\mathcal{L}_g}}{\partial(\partial_{\rho} 
h_{\alpha\beta})} - 2 \partial_{\lambda}
\left(\frac{\partial \bar{\mathcal{L}_g}}{\partial(\partial_{\rho}
\partial_{\lambda} h_{\alpha\beta})}\right) \right]
\left(\partial_{\rho}h_{\alpha\beta} 
- \frac{1}{4} \eta_{\alpha\beta}\partial_{\rho}h \right) 
\nonumber \\ &&
+ \, \frac{\partial \bar{\mathcal{L}_g}}{\partial(\partial_{\mu} \sigma)} 
\left(\partial_{\nu}\sigma + \frac{1}{8} \varphi_0
\partial_{\nu}h\right) -\frac{1}{4}\delta^{\mu}\,\!_{\nu}\frac{\partial 
\bar{\mathcal{L}_g}}{\partial(\partial_{\rho} \sigma)} 
\left(\partial_{\rho}\sigma + \frac{1}{8} \varphi_0
\partial_{\rho}h\right) \bigg\rangle, \nonumber \\ &&
\label{34}
\end{eqnarray}
where the angle brackets denote the average over a region of spacetime.
Next, we will use (\ref{34}) to calculate the energy and 
momentum carried by the plane waves of the three main
conformally invariant theories of gravity.


\subsection{Conformal dilaton gravity}
\label{sec4}


Let us start by calculating the linearized gravitational energy-momentum 
pseudo-tensor of the conformal dilaton gravity (CDG), whose Lagrangian density 
is given by \cite{Alv}
\begin{equation}
\mathcal{L}_{\textrm{CDG}} = \sqrt{-g}\left(\varphi^{2}R 
+ 6g^{\mu\nu}\partial_{\mu}\varphi\partial_{\nu}\varphi \right),
\label{35}
\end{equation}
where $R = g^{\mu\nu}R_{\mu\nu}$ is the scalar curvature, with $R_{\mu\nu} = 
R^{\alpha}\,\!\!_{\mu\alpha\nu}$ being the Ricci tensor, 
$R^{\alpha}\,\!\!_{\mu\beta\nu} = \partial_{\beta}\Gamma^{\alpha}_{\mu\nu} 
+ \cdots$ the Riemann tensor and $\Gamma^{\alpha}_{\mu\nu}$ the Levi-Civita 
connection.

Inserting (\ref{35}) into the field equations (\ref{20}) and (\ref{21}) 
in vacuum ($T_{\mu\nu} = 0$), we obtain the CDG field equations
\begin{equation}
\varphi^{2}G_{\mu\nu} +  6 \partial_{\mu}\varphi\partial_{\nu}\varphi 
- 3g_{\mu\nu}\partial^{\rho}\varphi\partial_{\rho}\varphi + g_{\mu\nu} 
\Box \varphi^{2} 
- \nabla_{\mu}\nabla_{\nu} \varphi^{2}  = 0,
\label{36}
\end{equation}
\begin{equation}
\left(\Box - \frac{1}{6}R \right) \varphi = 0,
\label{37}
\end{equation}
where
\begin{equation}
G_{\mu\nu} = R_{\mu\nu} - \frac{1}{2}g_{\mu\nu}R
\label{38}
\end{equation}
is the Einstein tensor and $\Box = \nabla^{\mu}\nabla_{\mu}$ is the 
generally covariant d'Alembertian operator.

Using the flat background field expansions (\ref{31}) and (\ref{32}) 
in (\ref{35})-(\ref{37}), we 
find the linearized CDG Lagrangian density
\begin{equation}
\bar{\mathcal{L}}_{\textrm{CDG}} =  \varphi_0^2\bar{\mathcal{L}}_{EH} + 
2\varphi_{0}\sigma\bar{R} + 6 \partial^{\mu}\sigma\partial_{\mu}\sigma,
\label{39}
\end{equation}
and the linearized CDG field equations
\begin{equation}
\varphi_{0}\left(\bar{R}_{\mu\nu} - \frac{1}{2}\eta_{\mu\nu}
\bar{R}\right) + 2 \eta_{\mu\nu}\bar{\Box}\sigma 
-2\partial_{\mu}\partial_{\nu}\sigma = 0,
\label{40}
\end{equation}
\begin{equation}
\bar{\Box}\sigma - \frac{\varphi_0}{6}\bar{R} = 0,
\label{41}
\end{equation}
where
\begin{equation}
\bar{\mathcal{L}}_{EH} =  \frac{1}{4} \Big( \partial^{\rho}
h^{\mu\nu}\partial_{\rho}h_{\mu\nu} - 2\partial^{\mu}h_{\mu\nu}
\partial_{\rho}h^{\rho\nu}+ 2\partial^{\mu}h_{\mu\nu}\partial^{\nu}h 
- \partial^{\mu}h\partial_{\mu}h  \Big)
\label{42}
\end{equation}
is the linearized Einstein-Hilbert Lagrangian density,
\begin{equation}
\bar{R}_{\mu\nu} = \frac{1}{2} \left( \partial_{\mu}\partial^{\rho}
h_{\rho\nu} + \partial_{\nu}\partial^{\rho}h_{\rho\mu} 
- \bar{\Box}h_{\mu\nu} - \partial_{\mu}\partial_{\nu}h  \right)
\label{43}
\end{equation}
is the linearized Ricci tensor,
\begin{equation}
\bar{R} =  \partial^{\mu}\partial^{\nu}h_{\mu\nu} 
- \bar{\Box}h 
\label{44}
\end{equation} 
is the linearized scalar curvature, and 
$\bar{\Box} = \partial^{\mu}\partial_{\mu}$.

Both the linearized Lagrangian density (\ref{39}) and the linearized 
field equations (\ref{40}) and (\ref{41}) are invariant under the 
coordinate gauge transformation
\begin{equation}
h_{\mu\nu} \rightarrow h_{\mu\nu} + \partial_{\mu}\xi_{\nu} + 
\partial_{\nu}\xi_{\mu},
\label{45}
\end{equation}
where $\xi^{\mu}$ is an arbitrary spacetime dependent vector field, and
under the conformal gauge transformations
\begin{equation}
h_{\mu\nu} \rightarrow h_{\mu\nu} + 2\eta_{\mu\nu}\Lambda,
\label{46}
\end{equation}
\begin{equation}
\sigma \rightarrow \sigma - \Lambda,
\label{47}
\end{equation}
where $\Lambda$ is an arbitrary spacetime dependent scalar field.

We may impose the Lorenz (harmonic) gauge
\begin{equation}
\partial^{\mu}h_{\mu\nu} - \frac{1}{2}\partial_{\nu}h = 0,
\label{48}
\end{equation}
which fixes the coordinate gauge freedom up to a residual coordinate gauge 
parameter satisfying
\begin{equation}
\bar{\Box}\xi_{\mu} = 0,
\label{49}
\end{equation}
and the unitary gauge
\begin{equation}
\sigma = 0,
\label{50}
\end{equation}
which fixes the conformal gauge freedom.

Using (\ref{48}) and (\ref{50}) in (\ref{39})-(\ref{41}), 
we obtain the gauge fixed Lagrangian density
\begin{equation}
\bar{\mathcal{L}}_{\textrm{CDG}} =  \frac{\varphi_0^2}{4}\left( \partial^{\rho}
h^{\mu\nu}\partial_{\rho}h_{\mu\nu} - \frac{1}{2}\partial^{\rho}h
\partial_{\rho}h \right),
\label{51}
\end{equation}
and the gauge fixed field equations
\begin{equation}
\bar{\Box}h_{\mu\nu} = 0, \ \ \ \ \ \ \ \bar{\Box}h = 0.
\label{52}
\end{equation}
Finally, substituting (\ref{51}) into (\ref{34}), we arrive at
\begin{equation}
\left\langle \bar{t}^{\mu\nu}\right\rangle = \frac{c\varphi_0^2}{2} \left\langle 
\partial^{\mu}h^{\alpha\beta} \partial^{\nu}h_{\alpha\beta} 
- \frac{1}{4} \partial^{\mu}h \partial^{\nu}h
- \frac{1}{4} \eta^{\mu\nu} \left(\partial^{\rho}h^{\alpha\beta} 
\partial_{\rho}h_{\alpha\beta} - \frac{1}{4} \partial^{\rho}h 
\partial_{\rho}h \right)\right\rangle.
\label{53}
\end{equation}
By taking the trace of (\ref{53}), we can directly see that 
$\left\langle \bar{t}^{\mu}\,\!_{\mu}\right\rangle = 0$, as expected.

It is well known that the plane-wave solution to (\ref{48}) and (\ref{52}) 
is given by
\begin{equation}
h_{\mu\nu} = a_{\mu\nu}\cos(k_{\rho}x^{\rho}),
\label{54}
\end{equation}
where $a_{\mu\nu}$ is a symmetric wave polarization tensor
and $k_{\mu}$ is the wave vector, which satisfy 
\begin{eqnarray}
k^{\mu}k_{\mu} &=& 0,
\label{55} \\
k^{\mu}a_{\mu\nu} &=& \frac{1}{2}k_{\nu}a,
\label{56}
\end{eqnarray}
with $a = \eta^{\mu\nu}a_{\mu\nu}$. The condition (\ref{56}) reduces the 
independent components of $a_{\mu\nu}$ from ten to six. In addition, we 
can chose a solution to (\ref{49}) to impose the four additional gauge 
conditions
\begin{equation}
a_{0i} = 0, \qquad a = 0,
\label{57}
\end{equation} 
which reduce the independent components of 
$a_{\mu\nu}$ to the same two as those of GR, where the index $i$ runs 
from $1$ to $3$.

Substituting (\ref{54}) into (\ref{53}), and using (\ref{57}),
we obtain
\begin{equation}
\left\langle \bar{t}^{\mu\nu}\right\rangle = \frac{c\varphi_0^2}{4}
\, a^{\alpha\beta}a_{\alpha\beta}\left(k^{\mu}k^{\nu}\right),
\label{58}
\end{equation}
which is equal the GR energy-momentum pseudo-tensor for 
$\varphi_{0}^{2} = c^3/16\pi G$. In order to find the value of $\varphi_0$, 
we consider the linearized field equation (\ref{52}) sourced by a flat 
conformal matter energy-momentum tensor, which is given by
\begin{equation}
-\frac{\varphi_0^2}{2}\bar{\Box}h_{\mu\nu} = \frac{1}{2c}\bar{T}_{\mu\nu},
\label{59}
\end{equation}
where we used (\ref{20}). 

Considering a conformal point particle source 
with mass $M$ at rest at the origin, for which \cite{Far1}
\begin{equation}
\bar{T}_{\mu\nu} = Mc^2\left(\delta^{0}_{\mu}\delta^{0}_{\nu} 
+\frac{1}{4}\eta_{\mu\nu}\right)\delta^{3}(\textbf{r}),
\label{60}
\end{equation}
and taking the Newtonian limit, we find that the 00 component of (\ref{59}) 
becomes
\begin{equation}
\nabla^2 \phi = \frac{3}{8} \frac{c^3 M}{\varphi_0^2} \delta^{3}(\textbf{r}),
\label{61}
\end{equation}
where $\phi = - c^2 h_{00}/2$ is the gravitational potential, $\nabla^2$ is 
the Laplacian operator and $r = |\textbf{r}|$ is the distance from the source.

For the solution to (\ref{61}), which is given by
\begin{equation}
\phi(r) = -\frac{3c^3 M}{32 \pi\varphi_0^2 r},
\label{62}
\end{equation}
to be consistent with the Newtonian potential
\begin{equation}
\phi(r) = - \frac{G M}{ r},
\label{63}
\end{equation}
we must have
\begin{equation}
\varphi_{0}^{2} = \frac{3c^3}{32\pi G}.
\label{64}
\end{equation}
This means that the magnitudes of the energy and momentum carried by the CDG 
plane waves are $1.5$ times bigger than those of the GR plane waves.


\subsection{Conformal gravity}
\label{sec5}


The conformal gravity (CG) is the only pure metric conformally invariant 
theory of gravity. Its Lagrangian density is given by \cite{Man1}
\begin{equation}
\mathcal{L}_{\textrm{CG}} = - \frac{1}{2\alpha^2}\sqrt{-g}\left( 
C^{\alpha\beta\mu\nu}C_{\alpha\beta\mu\nu}\right),
\label{65}
\end{equation} 
where $\alpha$ is a coupling constant, and
\begin{equation}
C^{\alpha\beta\mu\nu}C_{\alpha\beta\mu\nu} = R^{\alpha\beta\mu\nu}
R_{\alpha\beta\mu\nu} - 4R^{\mu\nu}R_{\mu\nu} + R^2
+ 2\left(R^{\mu\nu}R_{\mu\nu} - \frac{1}{3}R^{2}\right)
\label{66}
\end{equation}
is the Weyl tensor squared. 

Since the Euler density $\sqrt{-g}E = \sqrt{-g}\left(R^{\alpha\beta\mu\nu}
R_{\alpha\beta\mu\nu} - 4R^{\mu\nu}R_{\mu\nu} + R^2\right)$ does not contribute 
to the field equations of the theory, we can disregard it and write (\ref{65}) 
in the simplest form
\begin{equation}
\mathcal{L}_{\textrm{CG}} = - \frac{1}{\alpha^2}\sqrt{-g}\left( 
R^{\mu\nu}R_{\mu\nu} - \frac{1}{3}R^{2}\right).
\label{67}
\end{equation} 
The substitution of (\ref{67}) into the field equation (\ref{20}) in vacuum
gives the CG field equation
\begin{equation}
B_{\mu\nu} = 0,
\label{68}
\end{equation}
where
\begin{eqnarray}
B_{\mu\nu} &=& \Box R_{\mu\nu} 
- \frac{1}{3}\nabla_{\mu}\nabla_{\nu}R  -\frac{1}{6}g_{\mu\nu}\Box R  
+ 2R^{\rho\sigma}R_{\mu\rho\nu\sigma} 
-\frac{1}{2}g_{\mu\nu}R^{\rho\sigma}R_{\rho\sigma}  \nonumber \\ &&
- \frac{2}{3}RR_{\mu\nu}  + \frac{1}{6}g_{\mu\nu}R^2
\label{69}
\end{eqnarray}
is the Bach tensor.

Using (\ref{31}) and (\ref{32}) in (\ref{67}) and (\ref{68}), we 
find the linearized CG Lagrangian density
\begin{equation}
\bar{\mathcal{L}}_{\textrm{CG}} =  - \frac{1}{\alpha^2}\left( 
\bar{R}^{\mu\nu}\bar{R}_{\mu\nu} - \frac{1}{3}\bar{R}^{2}\right),
\label{70}
\end{equation}
and the linearized CG field equation
\begin{equation}
\bar{\Box} \bar{R}_{\mu\nu} 
- \frac{1}{3}\partial_{\mu}\partial_{\nu}\bar{R} 
- \frac{1}{6}\eta_{\mu\nu}\bar{\Box} \bar{R}  = 0.
\label{71}
\end{equation}
Both (\ref{70}) and (\ref{71}) are invariant under the coordinate gauge 
transformation (\ref{45}) and the conformal gauge 
transformation (\ref{46}). We can fix the coordinate 
gauge freedom by imposing the gauge
\begin{equation}
\partial^{\mu}h_{\mu\nu} = 0,
\label{72}
\end{equation}
and the conformal gauge freedom by imposing the gauge
\begin{equation}
h = 0,
\label{73}
\end{equation}
with $\xi^{\mu}$ satisfying
\begin{equation}
\bar{\Box}\xi_{\mu} =0, \ \ \ \ \ \ \ \partial^{\mu}\xi_{\mu} = 0.
\label{74}
\end{equation}

Using (\ref{72}) and (\ref{73}) in (\ref{70}) and (\ref{71}), we obtain 
the gauge fixed Lagrangian density
\begin{equation}
\bar{\mathcal{L}}_{\textrm{CG}} = - \frac{1}{4\alpha^2}\left(\bar{\Box}
h^{\mu\nu}\bar{\Box}h_{\mu\nu}\right),
\label{75}
\end{equation}
and the gauge fixed field equation
\begin{equation}
\bar{\Box}^2 h_{\mu\nu} = 0.
\label{76}
\end{equation}
The substitution of (\ref{75}) into (\ref{34}) then gives
\begin{equation}
\left\langle \bar{t}^{\mu\nu}\right\rangle = \frac{c}{\alpha^2}\left\langle 
\partial^{\mu}\partial^{\nu}h^{\alpha\beta}\bar{\Box}h_{\alpha\beta} 
- \frac{1}{4}\eta^{\mu\nu}\bar{\Box}h^{\alpha\beta}\bar{\Box}h_{\alpha\beta}
\right\rangle,
\label{77}
\end{equation}
where we used (\ref{72}) and (\ref{73}). It is not 
difficult to see that (\ref{77}) is traceless.

The plane wave solution to (\ref{72}), (\ref{73}) and (\ref{76}) is given by
\begin{equation}
h_{\mu\nu} = \left(a_{\mu\nu} + b_{\mu\nu} \eta_{\lambda}x^{\lambda}\right)
\cos(k_{\rho}x^{\rho}),
\label{78}
\end{equation}
where $a_{\mu\nu}$ and $b_{\mu\nu}$ are symmetric wave polarization tensors, 
$\eta_{\mu}$ is a time-like unit vector, and $k_{\mu}$ is a null wave vector, 
which satisfy
\begin{equation}
k^{\mu}k_{\mu} = 0, \ \ \ \ \ \ \ \eta^{\mu}\eta_{\mu} = -1,
\label{79}
\end{equation}
\begin{equation}
k^{\mu}a_{\mu\nu} = 0, \ \ \ \ \ \ \ a = 0,
\label{80}
\end{equation}
\begin{equation}
k^{\mu}b_{\mu\nu} = 0, 
\ \ \ \ \ \ \ \eta^{\mu}b_{\mu\nu} = 0, \ \ \ \ \ \ \ b = 0,
\label{81}
\end{equation}
with $b = \eta^{\mu\nu}b_{\mu\nu}$. The conditions (\ref{80}) and (\ref{81}) 
reduce the independent components of $a_{\mu\nu}$ to five and of $b_{\mu\nu}$ 
to two. We can choose a solution to (\ref{74}) to remove five more components 
from $a_{\mu\nu}$, meaning that the GR plane wave $a_{\mu\nu}\cos(k_{\rho}x^{\rho})$ 
has no physical degrees of freedom in CG  \cite{Fab}. Therefore, only two degrees of 
freedom of the plane wave $b_{\mu\nu} \eta_{\lambda}x^{\lambda}\cos(k_{\rho}x^{\rho})$ 
are physical in CG, which contradicts the well 
known result found in Ref. \cite{Rie} that the CG plane wave propagates six 
physical degrees of freedom. This discrepancy is due to the author of Ref. \cite{Rie}
not having correctly fixed the coordinate gauge freedom when working only with the 
traceless part of $h_{\mu\nu}$.

The substitution of (\ref{78}) into (\ref{77}) gives\footnote{It is worth noting 
that (\ref{82}) goes against the result found in Ref. \cite{Yan} that the energy 
carried by the conformal gravity plane wave diverges in momentum space. This is because 
the coordinate gauge freedom was also not correctly fixed in Ref. \cite{Yan}.}
\begin{equation}
\left\langle \bar{t}^{\mu\nu}\right\rangle = \frac{c}{\alpha^2} \,
b^{\alpha\beta}b_{\alpha\beta}\left(\eta^{\mu}k^{\nu} + \eta^{\nu}k^{\mu}
-\frac{1}{2}\eta^{\mu\nu}\eta_{\lambda}k^{\lambda}\right)\eta_{\rho}k^{\rho},
\label{82}
\end{equation}
which confirms that CG contains no plane gravitational waves of the type 
$a_{\mu\nu}\cos(k_{\rho}x^{\rho})$. Since  $\eta_{\rho}k^{\rho} < 0$, it follows 
from (\ref{82}) that the plane wave $b_{\mu\nu} \eta_{\lambda}
x^{\lambda}\cos(k_{\rho}x^{\rho})$ carries negative energy and momentum. However, 
this is not a problem, at least at the classical level, because there are no plane 
waves with positive energy and momentum in CG for the negative energy and momentum
plane wave to interact with, which prevents the energy and momentum conservation 
from being violated.

Just to finish, it is useful to analyze the Newtonian limit of the linearized 
field equation (\ref{76}) sourced by the conformal point particle 
energy-momentum tensor (\ref{60}), which is given by
\begin{equation}
\nabla^4 \phi = - \frac{3}{8}c^3 M \alpha^2 \delta^{3}(\textbf{r}).
\label{83}
\end{equation}
The fourth-order equation (\ref{83}) has as solution the repulsive potential 
that grows linearly with distance
\begin{equation}
\phi(r) = \left(\frac{3}{64\pi}c^3 M \alpha^2\right)r,
\label{84}
\end{equation}
which corresponds to the negative energy plane wave 
$b_{\mu\nu} \eta_{\lambda}x^{\lambda}\cos(k_{\rho}x^{\rho})$ that 
grows linearly with time. This is further proof that there is only the 
negative energy plane wave in CG. 


\subsection{Massive conformal gravity}
\label{sec6}


The Lagrangian density of the massive conformal gravity (MCG) has the form 
\cite{Far2}
\begin{equation}
\mathcal{L}_{\textrm{MCG}} = \sqrt{-g}\left[ \varphi^{2}R 
+ 6g^{\mu\nu}\partial_{\mu}\varphi\partial_{\nu}\varphi - \frac{1}{\alpha^2}
\left( R^{\mu\nu}R_{\mu\nu} - \frac{1}{3}R^{2}\right) \right], 
\label{85}
\end{equation}
where we have already disregarded the Euler density, for simplicity.

By substituting (\ref{85}) into the field equations (\ref{20}) and 
(\ref{21}) in vacuum, we find the MCG field equations 
\begin{equation}
\varphi^{2}G_{\mu\nu} +  6 \partial_{\mu}\varphi\partial_{\nu}\varphi 
- 3g_{\mu\nu}\partial^{\rho}\varphi\partial_{\rho}\varphi + g_{\mu\nu} 
 \Box \varphi^{2} 
- \nabla_{\mu}\nabla_{\nu} \varphi^{2} - \alpha^{-2} B_{\mu\nu} = 0,
\label{86}
\end{equation}
\begin{equation}
\left(\Box - \frac{1}{6}R \right) \varphi = 0.
\label{87}
\end{equation}
The use of (\ref{31}) and (\ref{32}) in (\ref{85})-(\ref{87}) gives 
the linearized MCG Lagrangian density
\begin{equation}
\bar{\mathcal{L}}_{\textrm{MCG}} =  \varphi_0^2\bar{\mathcal{L}}_{EH} + 
2\varphi_{0}\sigma\bar{R} + 6 \partial^{\mu}\sigma\partial_{\mu}\sigma 
- \frac{1}{\alpha^2}\left( \bar{R}^{\mu\nu}\bar{R}_{\mu\nu} 
- \frac{1}{3}\bar{R}^{2}\right),
\label{88}
\end{equation}
and the linearized MCG field equations
\begin{eqnarray}
&& \bar{\Box} 
\bar{R}_{\mu\nu} - \frac{1}{3}\partial_{\mu}\partial_{\nu}\bar{R} 
- \frac{1}{6}\eta_{\mu\nu}\bar{\Box} \bar{R}  \nonumber \\ &&
- \lambda^{-2} \left[ \bar{R}_{\mu\nu} - \frac{1}{2}\eta_{\mu\nu}
\bar{R} + \varphi_{0}^{-1}\left(2 \eta_{\mu\nu}\bar{\Box}\sigma 
-2\partial_{\mu}\partial_{\nu}\sigma\right)\right] = 0,
\label{89}
\end{eqnarray}
\begin{equation}
\bar{\Box}\sigma - \frac{\varphi_{0}}{6}\bar{R} = 0,
\label{90}
\end{equation}
where $\lambda = 1/(\varphi_{0}\alpha)$.

The linearized Lagrangian density (\ref{88}) and field equations (\ref{89})
and (\ref{90}) are invariant under the coordinate gauge 
transformation (\ref{45}) and the conformal gauge transformations (\ref{46}) 
and (\ref{47}). We can fix the coordinate gauge freedom by imposing the Lorentz 
gauge (\ref{48}) and the conformal gauge freedom by imposing the unitary gauge
(\ref{50}), which reduces (\ref{88}) to the gauge fixed Lagrangian density
\begin{equation}
\bar{\mathcal{L}}_{\textrm{MCG}} =  \frac{\varphi_0^2}{4}\left( \partial^{\rho}
h^{\mu\nu}\partial_{\rho}h_{\mu\nu} - \frac{1}{2}\partial^{\rho}h
\partial_{\rho}h \right)  
- \frac{1}{4\alpha^2}\left( \bar{\Box}
h^{\mu\nu}\bar{\Box}h_{\mu\nu} 
- \frac{1}{3}\bar{\Box}h\bar{\Box}h\right),
\label{91}
\end{equation}
and (\ref{89}) and (\ref{90}) to the gauge fixed field equations
\begin{equation}
\left(\bar{\Box} - \lambda^{-2} \right)\bar{\Box} h_{\mu\nu}  = 0, 
\ \ \ \ \ \ \ \bar{\Box} h  = 0.
\label{92}
\end{equation}

The plane wave solution to (\ref{92}) and (\ref{48}) is given by
\begin{equation}
h_{\mu\nu} = a_{\mu\nu}\cos(k_{\rho}x^{\rho}) 
+ b_{\mu\nu} \cos(q_{\rho}x^{\rho}),
\label{93}
\end{equation}
where $a_{\mu\nu}$ and $b_{\mu\nu}$ are symmetric wave polarization 
tensors, and $k_{\mu}$ and $q_{\mu}$ are wave vectors, which satisfy
\begin{equation}
k^{\mu}k_{\mu} = 0, \ \ \ \ \ \ \ q^{\mu}q_{\mu} = -\lambda^{-2},
\label{94}
\end{equation}
\begin{equation}
k^{\mu}a_{\mu\nu} = \frac{1}{2}k_{\nu}a, \ \ \ \ \ \ \ q^{\mu}b_{\mu\nu} 
= 0, \ \ \ \ \ \ \ b = 0.
\label{95}
\end{equation}
The conditions (\ref{57}) and (\ref{95}) reduce the independent components 
of $a_{\mu\nu}$ to two and of $b_{\mu\nu}$ to five.

Inserting (\ref{91}) into (\ref{34}), and using (\ref{92}), we obtain
\begin{eqnarray}
\left\langle \bar{t}^{\mu\nu}\right\rangle &=& \frac{c\varphi_0^2}{2}\bigg\langle  
\partial^{\mu}h^{\alpha\beta} \partial^{\nu}h_{\alpha\beta} 
- \frac{1}{4} \partial^{\mu}h \partial^{\nu}h
- \frac{1}{4} \eta^{\mu\nu}\partial^{\rho}h^{\alpha\beta} 
\partial_{\rho}h_{\alpha\beta}
\nonumber \\ && + \, 
2\lambda^{2}\left(\partial^{\mu}\partial^{\nu}h^{\alpha\beta}\Box 
h_{\alpha\beta} - \frac{1}{4}\eta^{\mu\nu}\bar{\Box}h^{\alpha\beta}
\bar{\Box}h_{\alpha\beta}\right) \bigg\rangle. 
\label{96}
\end{eqnarray}
The substitution of (\ref{93}) into (\ref{96}) then gives\footnote{This result, 
without the last term, was found before in Ref. \cite{Far3}. The difference 
between the two results is because the energy-momentum pseudo-tensor used in 
Ref. \cite{Far3} does not have the conformal symmetry as the one used here.}
\begin{equation}
\left\langle \bar{t}^{\mu\nu}\right\rangle = \frac{c\varphi_0^2}{4}
\left[a^{\alpha\beta}a_{\alpha\beta} \left(k^{\mu}k^{\nu}\right)
- b^{\alpha\beta}b_{\alpha\beta}\left(q^{\mu}q^{\nu} 
+ \frac{1}{4}\eta^{\mu\nu}\lambda^{-2}\right)\right],
\label{97}
\end{equation}
which can be shown to be traceless with the use of (\ref{94}).

We can see from (\ref{97}) that the massless plane wave 
$a_{\mu\nu}\cos(k_{\rho}x^{\rho})$ carries positive energy and momentum, while 
the massive plane wave $b_{\mu\nu}$ $\cos(q_{\rho}x^{\rho})$ 
carries negative energy and momentum. However, since these waves do not interact 
with each other, energy cannot flow between them so that there is 
no violation of energy and momentum conservation. At the quantum level, the 
negative energy field translates into a negative norm ghost state. However, the 
ghost state does not contribute to the gauge-invariant absorptive part
of the $S$-matrix in MCG \cite{Far4}, which leads to the unitarity of the theory.

The Newtonian limit of (\ref{92}) sourced by (\ref{60}) leads to the 
potential \cite{Far1}
\begin{equation}
\phi(r) = - \frac{GM}{r}
\left(1 - e^{-r/\lambda}\right),
\label{98}
\end{equation}
where we considered (\ref{64}). The potential (\ref{98}) is composed of an 
attractive Newtonian potential, which corresponds to the massless plane wave
$a_{\mu\nu}\cos(k_{\rho}x^{\rho})$, and of a repulsive Yukawa potential, which 
corresponds to the massive plane wave $b_{\mu\nu}\cos(q_{\rho}x^{\rho})$.
For (\ref{98}) to be consistent with solar system tests \cite{Far5} and 
Cavendish like experiments \cite{Adel}, we must have 
$\lambda \lesssim 3.3 \times 10^{-5} \, \mbox{m}$. 
Since the distance between the objects of a binary system in the inspiral phase 
is macroscopic, the contribution of the massive wave to the rate of energy 
loss from a binary system source is negligible. The contribution of the massless 
wave is $3/8$ of the GR value\footnote{Since the massive wave does not contribute 
to the rate of energy loss, the missing term in (\ref{97}) does not affect the 
result found in Ref. \cite{Far3}.} \cite{Far3}, which is within the precision of the 
measurements.


\section{Final remarks}
\label{sec7}


We have derived a gravitational energy-momentum pseudo-tensor that is 
manifestly covariant under conformal transformations. Then we used its 
average over a region of spacetime to find the energy and momentum 
carried by the gravitational waves of the three main conformally 
invariant theories of gravity, namely, conformal dilation gravity, conformal 
gravity and massive conformal gravity.

One of the results found is that the traceless condition of the conformal 
gravitational energy-momentum pseudo-tensor, which is a characteristic of all 
energy-momentum tensors with conformal symmetry, is independent of the gauge 
fixing. Individually, it was shown that the plane wave of conformal 
dilation gravity is identical of the general relativity plane wave but 
with a magnitude $1.5$ times bigger. In the case of conformal gravity, it was 
found that its plane wave has only two propagating physical degrees of freedom 
that grow with time and carry negative energy and momentum. Finally, the 
result found for massive conformal gravity is that its plane wave has seven 
non-interacting propagating physical degrees of freedom, with two of these 
degrees of freedom being massless and carrying positive energy and the other 
five being massive and carrying negative energy.

Since the results found here for plane waves in conformal gravity are different 
from those found in literature, it will be interesting  to use the gravitational 
energy-momentum pseudo-tensor developed here to study the energy carried by 
astrophysical gravitational waves in conformal gravity and see if the result 
agrees with the result found in Ref. \cite{Capr}.

Another interesting topic to be studied in the future is the calculation of the 
gravitational energy-momentum pseudo-tensor in conformally invariant theories 
of gravity formulated in the Palatini approach and checking whether it gives 
the same results for the gravitational energy density as the pseudo-tensor 
developed here in the metric approach. As such calculation leads to different 
results in other higher-order or scalar-tensor theories of gravity \cite{Abe}, it 
is likely that we can use it to discriminate between the metric 
and Palatini approaches of conformally invariant theories of gravity. Besides 
this, it will be useful to study in future works the role of non-locality in 
the pseudo-tensor developed here, since non-local corrections of the 
gravitational energy-momentum pseudo-tensor can give a relevant signature 
of quantum gravity effects \cite{Cap2,Cap3}.


\end{document}